\documentclass[usenatbib]{mn2e}
\usepackage{color} 

\usepackage{graphics}
\usepackage{epsfig}
\usepackage{natbib}

\begin{document}

\title[Luminous AGN in Massive Galaxies]
{Emission line properties of the most luminous AGN in massive galaxies at intermediate redshifts }

\author [G.Kauffmann \& C.Maraston] {Guinevere Kauffmann$^1$\thanks{E-mail: gamk@mpa-garching.mpg.de},
Claudia Maraston$^2$\\
$^1$Max-Planck Institut f\"{u}r Astrophysik, 85741 Garching, Germany\\
$^2$Institute of Cosmology and Gravitation, Dennis Sciama Building, Burnaby Road, Portsmouth PO1 3FX, UK}

\maketitle

%===================================
\begin{abstract} 
We have analyzed the emission line properties
of 6019 Type II AGN at redshifts between 0.4-0.8  with [OIII] luminosities
greater than $3 \times 10^8 L_{\odot}$, characteristic of the Type II
quasars first identified in population studies by Zakamska et al (2003).
The AGN are drawn from the  CMASS sample of galaxies with stellar masses
greater than $10^{11} M_{\odot}$ that were studied as part of the Baryon
Oscillation Spectroscopic Survey (BOSS) and comprise 0.5\% of the total
population of these galaxies. Individual spectra have low S/N,
so the analysis is carried out on stacked spectra in bins of [OIII]
luminosity and estimated stellar age.  The emission line ratios of the
stacks are well fit with simple uniform-density photoionization models
with metallicities between solar and twice solar.  In the stacks, a number
of emission lines are found to have distinct broad components requiring a
double Gaussian rather than a single Gaussian fit, indicative of outflowing
ionized gas.  These are: [OIII]$\lambda$4959, [OIII]$\lambda$5007,
[OII]$\lambda$3727,3729 and H$\alpha$$\lambda$6563.  Higher ionization
lines such as [NeIII]$\lambda$3869 and [NeV]$\lambda$3345 are detected in
the stacks, but are well fit by single Gaussians.  The broad components
typically contain a third of the total line flux and have widths of 600
km/s for the oxygen lines and 900  km/s for H$\alpha$.  The fraction of the
flux in the broad component and its width are independent of [OIII]
luminosity, stellar age, radio and mid-IR luminosity. The stellar mass of the galaxy
is the only parameter we could identify that influences the width of the broad line
component.

\end{abstract}


\section {Introduction}

Emission lines from ionized gas in active galactic nuclei (AGN) have long served
as probes of the physical conditions of gas in the vicinity of the actively
accreting black holes in these systems (see Osterbrock 1989 for a review). In Type I AGN , much of the emission
originates very close to the black hole from gas in the accretion disk (Shakura \& Sunyaev 1973). In Type
II AGN, the accretion disk is obscured by dust and the emission lines probe
highly ionized gas within the host galaxy (Antonucci 1993).

A standard analysis technique is to compare  observed emission line strengths
with photo-ionization models in order to constrain physical conditions within
the gas.  Baldwin, Phillips and Terlevich (1981) pointed out the importance of
the [OIII]$\lambda$5007/H$\beta$ versus [NII]$\lambda$6584/H$\alpha$ diagram
(hereafter the BPT diagram) for classifying emission-line galaxies as either
active galactic nuclei  or star-forming systems. In addition to the mechanism
responsible for the excitation of the gas, emission line ratio diagrams provide
important information about the spectrum of the ionizing source, as well as gas
temperature, density and metallicity. Line ratios also constrain the structure
of the gas.  e.g. whether it is in the form of radiation pressure dominated
dusty clouds (Groves et al 2004a,b) or matter and ionization-bounded clouds
(Binette et al 1996).

Emission line shapes provide additional important information about the
kinematics of the ionized gas in the galaxy. A large body of recent work has
focused on signatures of outflowing gas in powerful AGN, both at intermediate
and at high redshifts.  The most detailed of these studies are based on Integral
Field Unit (IFU) spectroscopy.  In a series of  papers, Liu et al (2013a,b)
studied  ionized gas nebulae in samples of around a dozen luminous, obscured,
radio-quiet quasars at $z \sim 0.5$ by means of Gemini Integral Field Unit
observations of the [OIII]$\lambda$5007 \AA\ line.  The nebulae were found to
have round morphologies and a mean diameter of 28 kpc. At each position within
the nebulae, the velocity structure of the [OIII]$\lambda$5007 \AA\ line was
found to be complex. The complicated line shapes were hypothesized to reflect
the existence of multiple gas components whose three-dimensional velocities are
projected onto the line-of-sight. Liu et al (2014) showed that the line shapes
could be well described by a superposition of 2-3 Gaussians.  
In some of the lines-of-sight through the nebulae, 
there was a blueshifted excess in the line profile and in others, there
was a redshifted excess, signifying that a signifigant fraction of the 
gas is outflowing, rather than in a
pressure-supported spherically symmetric cloud of halo gas  around the central nucleus.
The observed
velocity widths were used to estimate a median outflow velocity of 760 km/s.

Multi-Gaussian component fitting has by now become a standard technique to study
outflowing ionized gas in galaxies and has been applied extensively to samples
of both star-forming and active galaxies of different types (Genzel et al (2011)
(star-forming clumps in z-2 galaxies); Harrison et al 2017 (x-ray AGN); Talia et
al 2017; Davies et al 2019 (high-z star-forming galaxies); Toba et al 2017 (IR
bright dust-obscured galaxies); Schmidt et al (2018)(narrow-line Seyfert I);
Zakamska et al (2016) (high-z red quasars); Rose et al 2018 (ULIRGS).

As the samples under study have increased in size, comparative studies between
the emission line properties of different types of objects, studies of 
systematic trends in outflow properties as a function of star formation rate or
AGN luminosity, and comparisons of outflow signatures traced by a variety of
different gas tracers have become possible.  Some of the main findings so far
are that at fixed luminosity,  obscured and unobscured AGN have very similar
ionized gas nebulae sizes and kinematics (Liu et al 2014; Harrison et al 2018).
At fixed redshift, high outflow velocities are typical of AGN, but are rarely
found in star-forming galaxies (Harrison et al (2018). Forster-Schreiber et al
(2019)) find that the faster AGN-driven outflows become dominant in host
galaxies with stellar masses $\log M_* > 10.7$ and are found in 75\% of all
emission line galaxies with $\log M_* > 11.2$. Outflows traced by
ionized/neutral gas and by molecular gas have now been compared in a sample of a
dozen galaxies by Fluetsch et al (2019). In star forming galaxies, the ionized
outflow is estimated to carry as much total mass as the molecular outflow, but
in AGN, the molecular component dominates (see also a recent high-resolution
comparison of molecular and H$\alpha$ outflow components in a massive galaxy at
z=2 by Herrera-Camus et al (2018)).

In this paper we have carried out a systematic study of the emission lines in
the most luminous AGN found in the CMASS sample of galaxies observed as part of
the SDSS-III survey (Padmanabhan et al. 2012). The parent sample consists of
around a million galaxies in the redshift range $0.4<z<0.8$ with median stellar
mass of around $3 \times 10^{11} M_{\odot}$. The majority of these galaxies are
thus massive ``red-and-dead'' galaxies, but the adopted colour cuts do not
exclude systems with ongoing star formation and emission lines. The S/N of
individual spectra is low, but [OIII] can be reliably detected at luminosities
greater than $ 3 \times 10^8 L_{\odot}$, which corresponds to the threshold
[OIII] luminosity for an AGN to be included in catalogues of type II quasar
systems (Zakamska et al 2003). Redshift templates for galaxies are constructed by 
performing a rest-frame principal-component analysis (PCA) of training 
samples of known redshift. The leading eigenspectra from the PCA 
results are used to define a linear template basis that is used to model the 
spectra in the redshift analysis.  Statistical redshift errors propagated
from photon noise are typically a few tens of km/s for galaxies
(Bolton et al 2012). We adopt these pipeline redshifts as the systemic velocity of
the AGN in the analysis that follows.

 In recent work, Kauffmann (2018) studied a sample of
nearby AGN with comparable luminosities and showed that unlike the much more
numerous low-luminosity systems, the host galaxies of these object exhibited
clear signatures of recent starbursts and mergers/interactions. Although few in
number, they are responsible for the bulk of the recent black hole growth in the
most massive galaxies. In this work, by stacking spectra of comparable higher redshift systems in
bins of [OIII] luminosity and estimated mean stellar age, we recover very high
S/N composite AGN spectra and study trends in emission line ratios and shapes as
a function of these two quantities. We can also select subsamples that are
detected in the Faint Images of the Radio Sky at Twenty cm (FIRST) survey
(Becker, White \& Helfand  1995) or Wide-field Infrared Survey Explorer (WISE) survey (Wright
et al 2010) to investigate whether radio-loud, dusty sources differ
systematically in their emission line properties. In this way, we hope to gain
insight into the physical mechanisms responsible for signatures of outflowing
gas.

Our paper is structured as follows. In section 2, we describe our sample
selection, stacking and line-fitting procedure. In section 3, we show  how
emission line ratios vary as a function of [OIII] luminosity and mean stellar
age and we compare our derived sequences of emission line ratios to available grids
of photo-ionization models by Groves et al (2004). In section 4, we carry out an
examination of emission lines in a small sample of very luminous AGN in the
nearby Universe. In section 5, we move onto an analysis of the emission lines
shapes in the  composite spectra as quantified by multi-Gaussian line fitting.
In section 6, we conclude by summarizing our results and discussing future
perspectives.

\section {Sample selection and line-fitting}

\subsection{Sample construction} Our parent sample in constructed from the Wisconsin
catalogue of principal component analysis (PCA)-based stellar masses and
velocity dispersions that has been publicly released for the data release 12
(DR12) of the SDSS (Alam et al 2015). The methodology for deriving stellar
masses and other physical parameters from the spectra is described in detail in
Chen et al (2012). In brief, the steps in the method are as follows.  A library
of model star formation histories is generated with a range in parameters such
as the e-folding time of SFR(t), the recent burst mass fraction, the metallicity
of the stellar population, and the amount of dust extinction.  Principal
components (PCs) are identified from the model library after subtraction of the
mean spectrum calculated from the entire library. Physical parameters are
estimated after projecting the observed spectrum onto the PCs. A weight $w_i =
\exp(-\chi_i^2/2)$ is defined to describe the similarity between the given galaxy
and model $i$ and a  probability distribution function (PDF) is then built for
each parameter P, by looping over all the model galaxies in the library and by
summing the weights $w_i$ at the value of P for each model.  A set of 7 PCs are
provided in the Wisconsin catalogue, from which it is possible to reconstruct a
variety of physical parameters (see Table 1 of Chen et al (2012)).  In this
paper, we make use of the stellar mass and the estimated mean stellar age of
the stellar population. We note that the model-fitting
assumes the   universal  initial  mass  function  (IMF)  given in
Kroupa  (2001).

We select all galaxies from the catalogue in the redshift range $0.4<z<0.75$,
yielding a sample of 1,094,697 galaxies. Almost all of these galaxies are
included in the ``CMASS'' sample, so-named because it is very approximately 
stellar-mass limited even though it is selected by colour (Padmanabhan et al 2012).
These are then cross-correlated with the
Portsmouth catalogue of stellar kinematics and emission line fluxes (Thomas et
al 2013), which uses adaptations of the publicly available codes Penalized PiXel
Fitting (pPXF, Cappellari \& Emsellem 2004) and Gas and Absorption Line Fitting
code (GANDALF v1.5; Sarzi et al. 2006) to calculate stellar kinematics and to
derive emission line properties. We find 6019 galaxies with [OIII]$\lambda$5007
line luminosities greater than $3 \times 10^8 L_{\odot}$ (i.e 0.5\% of the parent
sample).  In 94\% of these systems, the [OIII] line is detected with $S/N > 3$
and in 78\% , it is detected with $S/N>5$, i.e.  the [OIII] line luminosity can
be measured in all these galaxies with reasonable accuracy.

Figure 1 shows histograms of stellar masses, redshift, mean stellar ages, and
[OIII] luminosities for the parent sample (solid black histograms) and
[OIII]-detected galaxies (red histograms) \footnote{In this plot we show all
galaxies with [OIII] detected down to a limiting line luminosity of $10^8
L_{\odot}$.} We have created sub-samples of [OIII]-detected objects with radio
source counterparts in the FIRST VLA survey (dashed red histograms) and with
detections in the W1 and W2 bands in the WISE source catalogue (dotted red
histograms). As can be seen, all of the samples have very similar distributions
of stellar masses. The [OIII]-detected samples are shifted to higher redshifts
and lower mean stellar ages compared to the parent sample. The shift to higher
redshift occurs because CMASS galaxies are selected a fixed limited magnitude in the 
$i$-band and the [OIII] line is redshifted through this bandpass over the redshift
interval spanned by the galaxies in this sample. The shift to younger ages is in
accord with the findings of Kauffmann et al (2003b) that luninous AGN have younger
stellar ages than control galaxies of the same stellar mass and morphological type. The radio and
WISE-detected subsamples do not differ significantly in [OIII] line luminosity
or mean stellar age compared to the full [OIII]-detected sample.

\begin{figure}
\includegraphics[width=90mm]{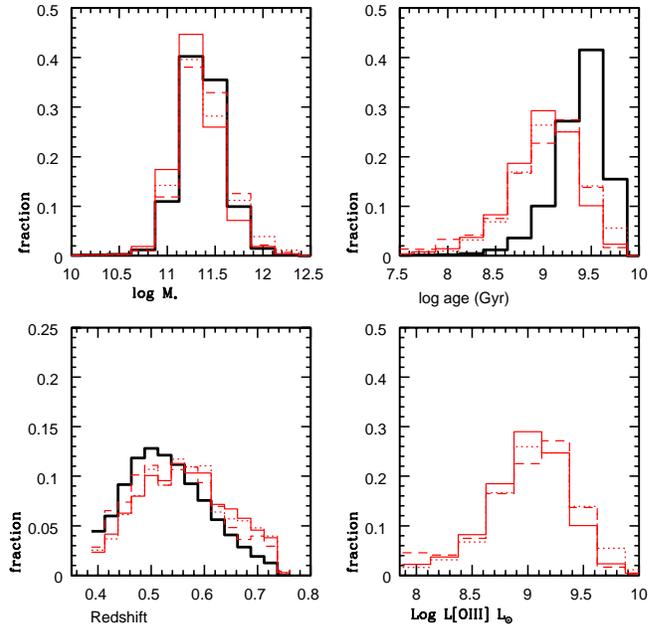}
\caption{ Histograms of the fraction of galaxies as a function of
stellar mass, redshift, mean stellar age, and
[OIII] luminosity for the parent sample (solid black histograms) and
for [OIII]-detected galaxies with [OIII] luminosities greater
than $10^8 L_{\odot}$ (solid red histograms). Results are also plotted for 
sub-samples of [OIII]-detected objects with radio
source counterparts in the FIRST VLA survey (dashed red histograms) and with
detectections in the W1 and W2 bands in the WISE source catalogue (dotted red
histograms).  
\label{models}}
\end{figure}

\subsection {Stacking procedure} We divide our sample of 6019 galaxies into 3
bins in [OIII] line luminosity: $\log$ L[OIII] = 8.5-9.0; 9.0-9.5; 9.5-10.0, and
4 bins in mean stellar age: $\log$ age(yr) = 8.0-8.4, 8.4-8.8, 8.8-9.2, and
9.2-9.6. Emission lines are usually quite weak in the oldest bin, so in some
cases we will only  show results for a total of 9 stacked spectra rather than
12. The number of galaxies in each bin range from $\sim 50$ in the highest
luminosity bins to $\sim 1500$ in the lower luminosity ones.

Our stacking methodology follows the procedure laid out in Vanden Berk et al
(2001), who created a variety of composite quasar spectra using a homogeneous
data set of over 2200 spectra from the Sloan Digital Sky Survey (SDSS).  We
first mask skylines and bad pixels in each spectrum and interpolate the flux
over the masked pixels. The spectra are rebinned onto a 1 \AA\ linear grid in
rest-frame wavelength. The spectra are ordered in redshift from lowest to
highest. The stacking then proceeds as follows. The rest-frame wavelength
boundaries of spectrum(N=i) are determined and the mean flux is calculated over
these boundaries. The mean flux of spectra(N=1...i-1) is then computed over the
same wavelength boundaries and the mean flux spectrum(n=i) is rescaled by a constant
factor such that they match. In this way, every spectrum in the stack
carries equal weight, regardless of the redshift or the luminosity of the
source.  We have calculated both mean and median stacks of the flux and find
that emission lines are stronger in the mean stacks, so these are
utilized in this paper. As we will show, the emission line shapes do
not depend strongly on the luminosity of the AGN in the stack, so
using mean stacks will not bias our results. Finally, we boxcar smooth the spectra to a resolution of
3 \AA\ and calculate  the error on the mean spectrum by a boot-strap resampling
process where the stacking is repeated 100 times on randomly selected spectra drawn from 
the given bin  of [OIII] luminosity and mean stellar age.

\subsection {Continuum fitting} For continuum fitting, we have implemented the
algorithm described in Wilkinson et al (2017), who introduced a code called
FIREFLY (Fitting Iteratively for Likelihood analysis).  The method is based on
fitting combinations of single stellar populations (SSPs), following an
iterative, best-fitting process until convergence is reached based on a Bayesian
Information Criterion. The algorithm is described in detail in Section 3.2 of
Wilkinson et al, and so we will not provide a detailed description of all the
steps here.

In order to span as wide a wavelength range as possible in the spectral fitting,
we use the stellar population models of Maraston \& Stromback (2011), which have
been extended into the UV using the  theoretical spectral library UVBLUE
(Rodrigues-Merino et al. 2005). We note that the first version of the
UV-extended models were published in Maraston et al (2009) and included fully
theoretical, high-resolution SEDs for wavelengths  $\lambda < 4350$ \AA. The
optical part of the SSPs are constructed using the MILES stellar library
(S\'anchez-Blazquez et al 2006), which provides observed spectra for 1000 stars
covering most evolutionary stages.  In our fitting procedure, we utilize SSPs
that span a range in ages from 0.055 to 12 Gyr  at half solar, solar and twice
solar metallicities -- this should  cover the likely range of metallicities in
massive galaxies with $\log M_* \sim 11$. Once again, a Kroupa (2001) IMF
has been adopted.

The SSPs are convolved with a Gaussian function with $\sigma$=200 km/s to match
the typical velocity dispersion of the galaxies in our sample and then boxcar
smoothed to the same 3 \AA\ resolution as the real stacked spectra. Regions
around known emission lines are masked before carrying out the fitting. We
implement the two-parameter dust extinction model of Wild et al (2007) on each
SSP, which has the form: \begin {equation} \tau_{\lambda}/\tau_v= (1-\mu)
(\lambda/5500\AA)^{-1.3}+ \mu(\lambda/5500\AA)^{-0.7} \end{equation} 
where $\tau_V$ is the total effective optical depth in the V-band and
$\mu$ is the fraction of the total optical depth contributed by the ambient
interstellar medium. This form of extinction law is motivated by the dust model
of Charlot \& Fall (2000) and has been shown to be broadly consistent with
observational determinations (Wild et al 2011). We allow $\mu$ to span the full
range from 0 to 1 and $\tau_V$ to vary from 0 to 3. Note that each fitted  SSP is
allowed to have different values of $\tau_V$ and $\mu$. In practice we find that
our AGN spectra are best fit by combining old SSPs with low dust extinction with
young SSPs with high dust extinction, as would be expected if these systems are
undergoing or have undergone a recent burst of star formation.

An example of a continuum fit is shown in Figure 2 for the stack with $\log$
L[OIII]=8.5 and mean stellar age between $6.3 \times 10^8$ and $1.5 \times
10^9$ Gyr.  The stacked spectrum is plotted in black, with vertical black lines
delineating masked regions of the spectrum(the actual emission lines in the stacked spectrum
are shown in blue).  The cyan bands across the bottom of each of the panels
indicate the $\pm 1\sigma$ errors on the mean observed flux. The best fit model
spectrum is shown in red. Note that the rest-frame wavelength range over which
the SSP model fitting is carried out is  2500-7000 \AA\, and Figure 2 shows that there is a
good match to the continuum level across this entire interval.

\begin{figure*}
\includegraphics[width=140mm]{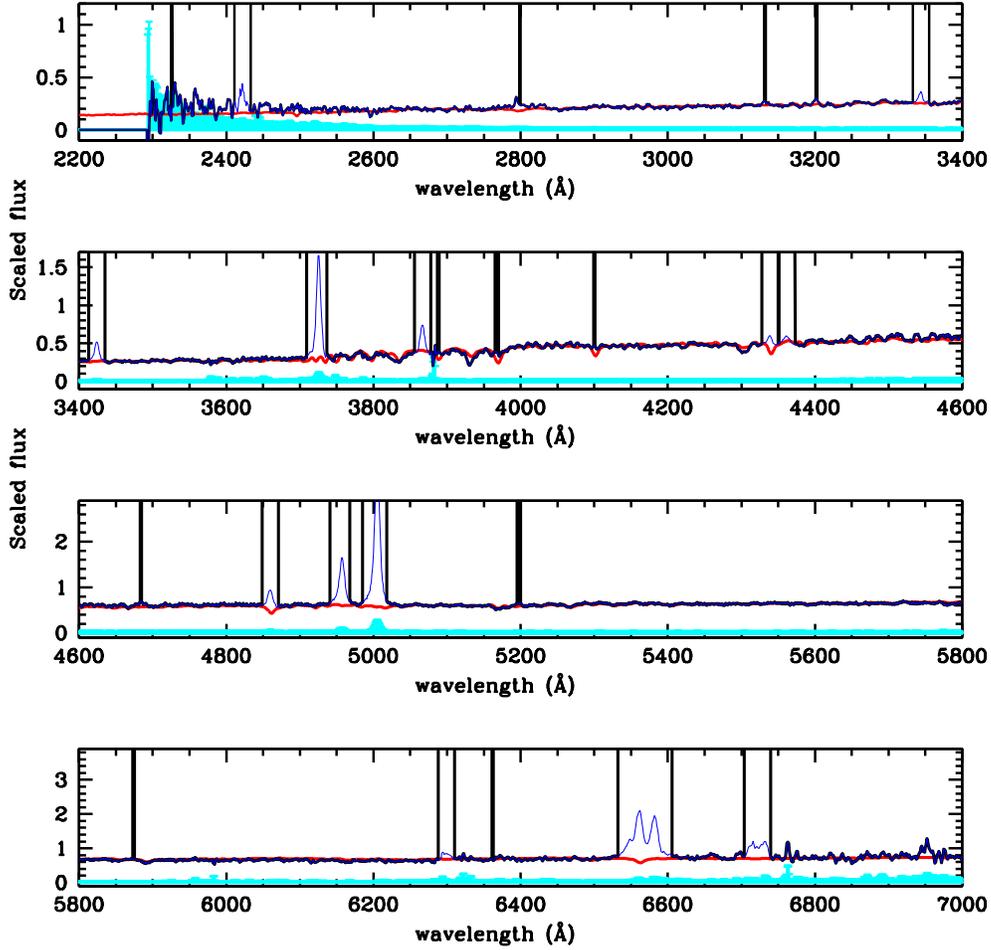}
\caption{ 
An example of a continuum fit is shown in for the stack with $\log$
L[OIII]=8.5 and mean stellar age between $6.3 \times 10^8$ and $1.5 \times
10^9$ Gyr.  The stacked spectrum is plotted in black, with vertical black lines
delineating masked regions (the actual emission lines in the stacked spectrum
are shown in blue).  The cyan bands across the bottom of each of the panels
indicate the $\pm 1\sigma$ errors on the mean observed flux. The best fit model
spectrum is shown in red. The emission lines that are analyzed in this paper
are marked. 
\label{models}}
\end{figure*}

\subsection {Line fitting} We carry out emission line fitting in a series of
iterative steps of increasing complexity. In order of increasing wavelength, the
emission lines that are generally well-detectable in most of the stacked spectra
are: [Mg II]$\lambda$2796,2803, HeII$\lambda$3203, [NeV]$\lambda$3345,
[OII]$\lambda$3726,3728, [NeIII]$\lambda$3869, H$\delta\lambda$4102,
H$\gamma$ $\lambda$4340, [OIII]$\lambda$4363, [HeII]$\lambda$4686,  H$\beta$$\lambda$4861,
[OIII]$\lambda$4959, [OIII]$\lambda$5007, [OI]$\lambda$6300,
H$\alpha$$\lambda$6563, [NII]$\lambda$6583.

The first step is to subtract the best-fit continuum, and to perform a fit to
the line using a single Gaussian, with line amplitude, width and centroid as
free parameters (the centroid can vary within $\pm 10$ \AA\ of the rest-frame
wavelength). Even after subtraction of the best-fit continuum,  
a small systematic offset from zero is sometimes found at
large wavelength separation from the line centroid, indicative of a systematic
error in the continuum estimate. These offsets are usually very small compared
to the peak flux of the line ($<1$\% in all cases). We allow for a constant
additive correction to the best-fit continuum,  refit the line and adopt this as
our corrected continuum if $\chi^2$ has decreased.  The next step is to perform
a double Gaussian fit.  In the first instance, we adopt the same centroid for
the secondary component as derived for the single Gaussian fit. If $\chi^2$
decreases by 30\% or more, then we record the secondary component as a clear
detection.  Finally, we refit the double Gaussian allowing the centroid of the
secondary component to vary to obtain final parameter values for both
components.

As we will show in Section 5, either one or two Gaussian components are sufficient to
fit nearly all the lines in the stack. In the case of two closely spaced lines,
line-fitting results from one line may need to be included when fitting 
the second one.  For example, in the case of
[OIII]$\lambda$4959 and [OIII]$\lambda$5007, we fit the weaker 4959 line first.
In almost all cases, a clear broad component is detected and careful subtraction
of its contribution redward of the line centroid is necessary before a fit to
[OIII]$\lambda$5007 can be performed.  We will illustrate the case of the
H$\alpha$/[NII] complex in more detail in Section 5.

\section{Emission line ratio trends as a function of [OIII] luminosity and age}

After fitting and subtracting the emission lines from the stacked spectrum, we
measure the age-sensitive stellar continuum indices  D$_n$(4000) (the narrow
version of the 4000 \AA\ break strength defined in Balogh et al (1999)) and the
H$\delta_A$ Lick index defined in Worthey \& Ottaviani (1997). In Figure 3, the
top two panels show the measured values of D$_n$(4000) and H$\delta_A$ as a
function of the logarithm of the mean stellar age. Each stellar age bin is
indicated by a filled square.  Results are shown in different colours for the 3
different bins in $\log$ L[OIII]: green ($\log$ L[OIII]=8.5-9); red ($\log$
L[OIII]=9-9.5); magenta ($\log$ L[OIII]=9.5-10).  The relation between
D$_n$(4000) and stellar age is independent of [OIII] luminosity, but H$\delta_A$
is considerably stronger at ages between a few $\times 10^8$ years and $10^9$
years in the more [OIII]-luminous stacks, which indicates that the more luminous
AGN are likely to be post-starburst systems. The bottom left panel shows H$\delta_A$
as a function of D$_n$(4000), where we see that AGN in the two more luminous
stacks exhibit the classic signature of enhanced Balmer absorption at fixed 4000
\AA\ break strength (Kauffmann et al 2003a). This finding is consistent with the
conclusions presented in Kauffmann (2018) showing that a large fraction of the
black hole growth in very luminous AGN takes place in galaxies that have
experienced a recent burst of star formation.

The bottom right panel of Figure 3 shows the Balmer decrement
(H$\alpha$/H$\beta$), which is a measure of the dust extinction in the narrow
line region,  as a function of the 4000 \AA\ break. Note that the dust-free
value for case B recombination is 2.86, so there is evidence of dust in these
systems. There is, however, very little variation in the measured Balmer
decrement with either [OIII] luminosity or with mean stellar age.

\begin{figure}
\includegraphics[width=83mm]{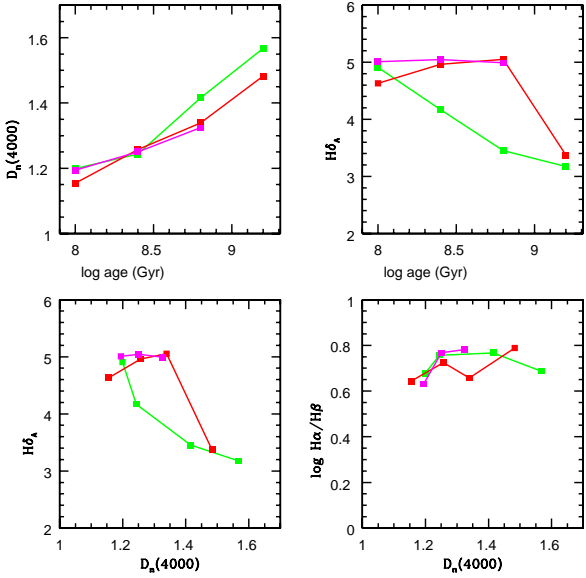}
\caption{ 
{\em Top:} measured values of D$_n$(4000) (left) and H$\delta_A$ (right) as a
function of the logarithm of the mean stellar age. 
{\em Bottom left:} H$\delta_A$
as a function of D$_n$(4000). {\em Bottom right:} Balmer decrement
(H$\alpha$/H$\beta$)
as a function of the 4000 \AA\ break.
Each stellar age bin is
indicated by a filled square.  Results are shown in different colours for the 3
different bins in $\log$ L[OIII]: green ($\log$ L[OIII]=8.5-9);
red ($\log$
L[OIII]=9-9.5); magenta ($\log$ L[OIII]=9.5-10). 
\label{models}}
\end{figure}

In Figure 4, we plot a number of ratios that are sensitive to the ionization
state of the gas and the spectral energy distribution of the ionizing radiation
field as a function of the mean stellar age of the galaxies  in the stack. Once
again, results are shown for 3  different ranges in [OIII] luminosity. The top
two panels show the ratios [OIII]$\lambda$5007/H$\beta$ and 
[OIII]$\lambda$5007/[OII]$\lambda$3727,3729. Because the stacks
are binned according [OIII] luminosity, the y-axis separation is to some extent
a selection effect. The main conclusion from the upper two panels is that the
ionization parameter does not appear to depend on mean stellar age at fixed
[OIII] luminosity. When combined with our result that the most luminous AGN tend to have
post-starburst stellar populations,  we find a clear result that radiation from the black hole rather
than from young stars is the dominant source of ionizing photons in these
systems. The bottom two panels show the ratios [Ne III]$\lambda$3869/[OII]$\lambda$3727,3729 
and [NeV]$\lambda$3345/[NeIII]$\lambda$3869
as a function of age. Here, the differences between the different
[OIII]-luminosity bins are much weaker, and are in the sense of the less
luminous AGN having slightly higher ionization or harder-spectrum radiation
fields. Once again, there is little or no dependence of these ratios on mean
stellar age (the [NeV/[NeIII] ratio shows a slight drop for stellar ages greater
than $10^9 M_{\odot}$.).

\begin{figure}
\includegraphics[width=83mm]{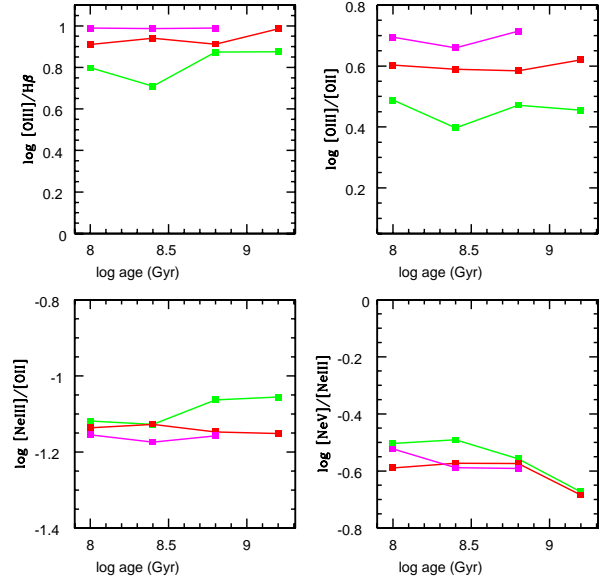}
\caption{A variety of line ratios sensitive to the ionization state of the gas
and the spectrum of the ionizing radiation are plotted as a function of
the mean age of the stellar population in the host galaxy.  
\label{models}}
\end{figure}

We now compare our derived emission line ratios to grids of photo-ionization
models provided by Groves, Dopita \& Sutherland (2004, hereafter GDS).  This
comparison is intended to act as a check on our line measurements rather serving
as a stringent  test of any physical model. As discussed in GDS, the very
simplest models assume a constant density of gas and allow for variations in
ionization parameter, metallicity and spectral shape of the ionizing radiation
field. Constant density models are doubtless oversimplified and take no account
for the gas-dynamics of the photo-ionized region. The next level of complexity
is to construct simple models with more realistic physics, but that necessarily
contain more assumptions.  Dopita et al (2002) developed a model for the NLR as
a dusty,radiation pressure dominated region surrounding a photo-evaporating
molecular cloud, which in turn is surrounded by a coronal halo of gas within
which the dust has been largely destroyed. They further assumed that the
emission spectrum is generated by the dusty dense inner region near the
stagnation point in the flow around the cloud. They argued that this would be
the highest emissivity region in the flow, and it would also be the region in
which the radiation pressure gradient is matched by the gas pressure gradient,
thus allowing for static rather than dynamic photoionization models.

Figure 5 compares derived emission line ratios from our stacked spectra with
constant density models. Once again, green, red and magenta squares correspond
to measurements from our 3 bins in [OIII] line luminosity. There is very little
dependence in observed line ratios with stellar age, except for the case of
[HeII]/H$\beta$, which we will discuss in more detail later. Cyan, blue and
dark gold open circles correspond to models with metallities half solar, solar and
twice solar, with ionization parameters ranging from U=0 to U=-3.  We adopt a
spectral slope $\alpha=-1.4$. The models of GDS span a range
in spectral slope between -1.2 and -2, and $\alpha=-1.4$ is their adopted
fiducial value. The measured
points lie fairly consistently in the region of model space between the solar
and twice solar models with ionization parameter U=-2 (the twice solar model
with U=-2 is marked as a blue asterisk in each panel). The only exception
are line ratios involving [HeII]$\lambda$4386.  Figure 6 is the
corresponding figure for the dusty model. In this case, we see that there is no
model that is able to fit the majority of the line ratios consistently.
Constraints from the [OIII]$\lambda$5007/H$\beta$ versus  [NII]$\lambda$6583/H$\alpha$
diagram and the [OIII]$\lambda$4363/[OIII]$\lambda$5007 versus
[NeV]$\lambda$3426/[NeIII]$\lambda$3866 diagram  would appear to favour models
with metallicities between solar and twice solar. In contrast, the
[OIII]$\lambda$5007/H$\beta$ versus  [OI]$\lambda$6300/H$\alpha$ diagram clearly
favours models with metallicities near half solar. The fact that we can find a single
metallicity that fits most line ratios for the  dust-free
models and we cannot do the same for the radiation-pressure
dominated models may, at first glance, lead us to favour
the dust-free models for these very luminous systems. We note, however, that the
Balmer decrement values shown in the lower right panel of Figure 3 clearly   
indicate the presence of dust, so the disagreement may just reflect short-comings
in the Dopita et al  model for the dynamics of the gas in the  narrow-line regions.  

\begin{figure*}
\includegraphics[width=145mm]{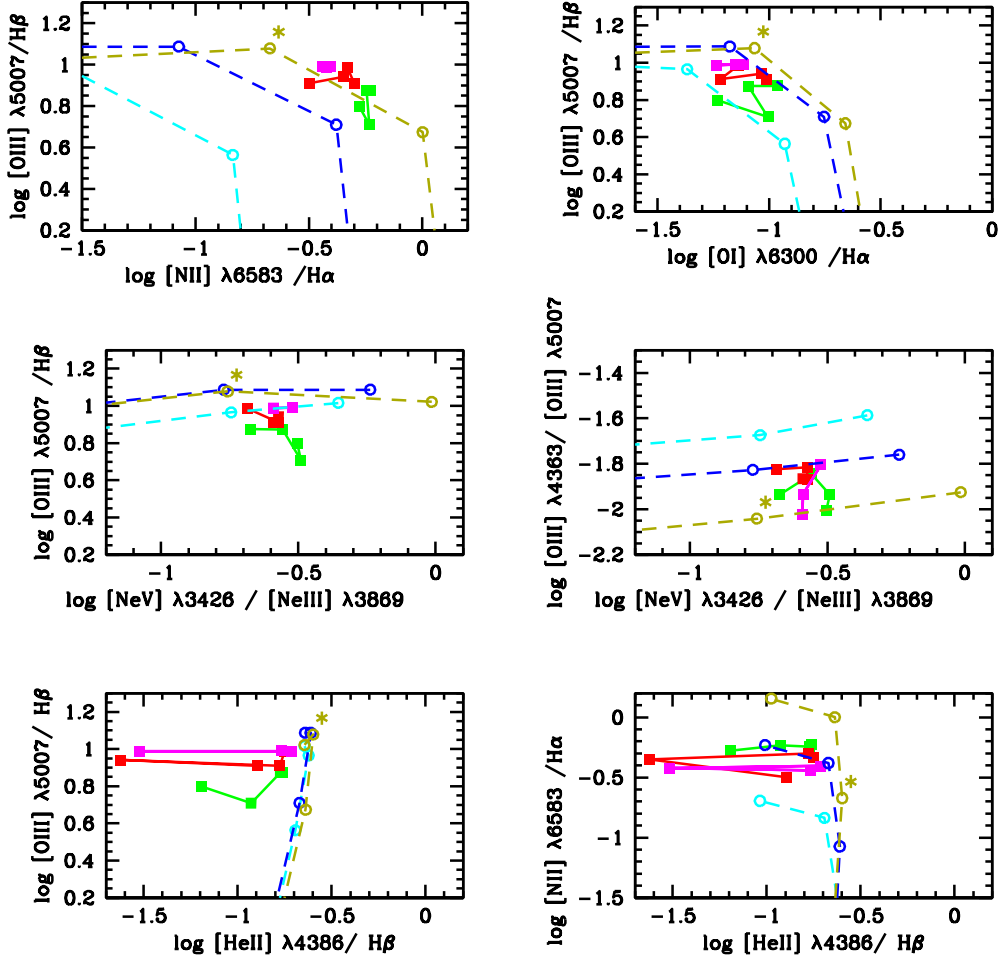}
\caption{ Emission line ratios from our stacked spectra are compared with dust-free uniform density 
photo-ionization  models from Groves et al (2014). Green, red and magenta squares correspond
to measurements from our 3 bins in [OIII] line luminosity. Cyan, blue and
dark gold open circles correspond to models with metallities half solar, solar and
twice solar, with ionization parameters ranging from U=0 to U=-3.  We adopt a
spectral slope $\alpha=-1.4$. The twice solar model
with U=-2 is marked as a dark gold  asterisk in each panel. 
\label{models}}
\end{figure*}

\begin{figure*}
\includegraphics[width=145mm]{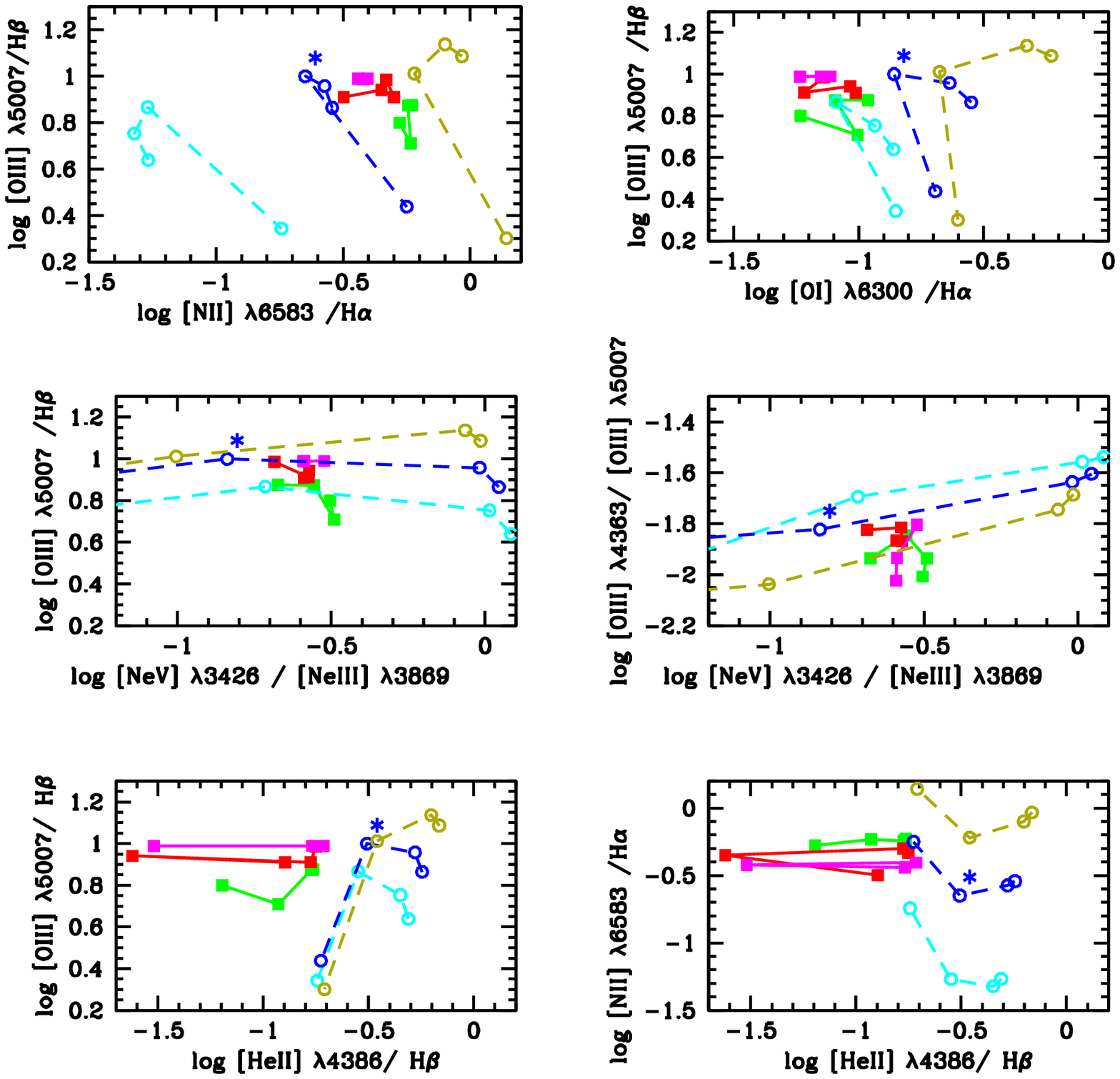}
\caption{ As in Figure 5, except for the dusty, radiation pressure dominated photoionization models.  
Here, the solar model with U=-2  is marked as a blue asterisk.
\label{models}}
\end{figure*}

Note, however, that both models fail to fit the range in [HeII]$\lambda$4386/H$\beta$ seen in the
stacked spectra. We examine this in more detail in Figure 7, where we plot
[HeII]$\lambda$ 4386$/H\beta$ as a function of stellar age. Here we see that the
pronounced ``dip'' in [HeII]$\lambda$ 4386/H$\beta$  at a stellar age of $2-5
\times 10^8$ yr in the two high luminosity bins is responsible for the offset
from the predictions of the two photo-ionization models.  This is the age
corresponding to a predominantly post-starburst stellar population in the host
galaxies of these AGN. In recent work attempting to reconcile the stellar and
nebular spectra of high-redshift galaxies, Steidel et al (2016) examine HeII
emission by means of the HeII$\lambda$1640/H$\beta$ ratio, pointing out that
massive star binaries produce a harder ionizing UV spectrum with a different
duty cycle than single-star models. We leave further examination of the observed
[HeII] anomaly in our AGN spectra for future work.

\begin{figure}
\includegraphics[width=70mm]{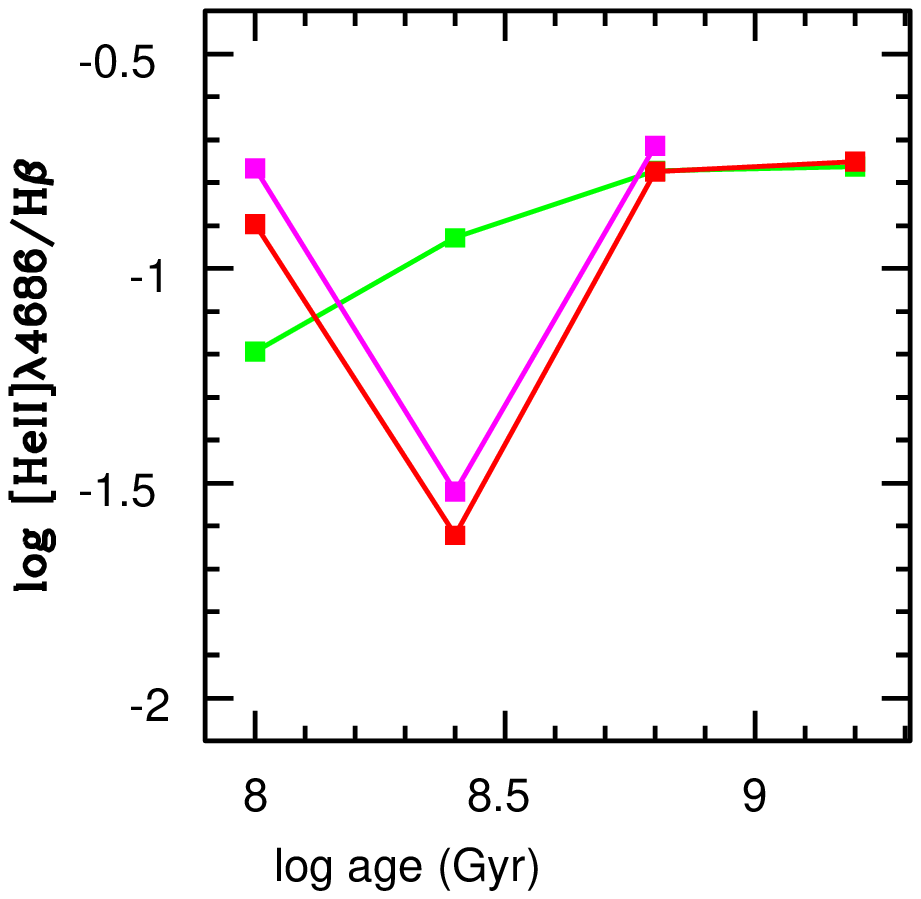}
\caption{ Measured values of [HeII]$\lambda$4386/H$\beta$ as a
function of the logarithm of the mean stellar age.   
Results are shown in different colours for the 3
different bins in $\log$ L[OIII]: green ($\log$ L[OIII]=8.5-9);
red ($\log$
L[OIII]=9-9.5); magenta ($\log$ L[OIII]=9.5-10).
\label{models}}
\end{figure}

\section {Examination of emission line shapes in nearby luminous AGN}

Before proceeding to the results for the stacked spectra, it is instructive to
examine high signal-to-noise individual spectra of nearby objects with similar
intrinsic luminosities. Stacking will always smooth out features: apparent
Gaussian components are likely only Gaussian in shape because we have combined
spectra from many galaxies. In this section, we examine some of the strongest
emission lines in 8 galaxies with [OIII] luminosities greater than $10^9
L_{\odot}$ drawn from the sample presented in Kauffmann (2018).

We find that the sources can be split into two main categories. The first
category is comprised of sources with clear outflow signatures. There are four
such sources and they are shown in Figure 8. The central narrow line component is of the
[OII] and [OIII] lines are generally blue-shifted with respect to the systemic
velocity (indicated as the dotted vertical line in each panel), and the line
shapes are asymmetric with  clear blue and/or red-wing excesses seen at the base
of the narrow-line component. A blue-side excess is usually also seen in the
H$\alpha$ line. The other four sources in our sample of 8 have double-peaked
narrow emission line profiles, but no clear outflow signatures (Figure 9).  We
note that all of the sources in the sample studied by Kauffmann (2018) are
detected in the FIRST VLA radio catalogue and have radio luminosities greater
than $10^{23.5}$ Watts Hz$^{-1}$. It is well-known that radio-loud AGN often
exhibit double-peaked narrow emission lines (e.g. Eracleous \& Halpern 2003).  A
variety of scenarios have been proposed to explain these features, including
supermassive binary black holes, emission from a bipolar outflow,
anisotropically illuminated broad line regions, and illumination of the outer
accretion disk by the inner accretion disk.

\begin{figure}
\includegraphics[width=90mm]{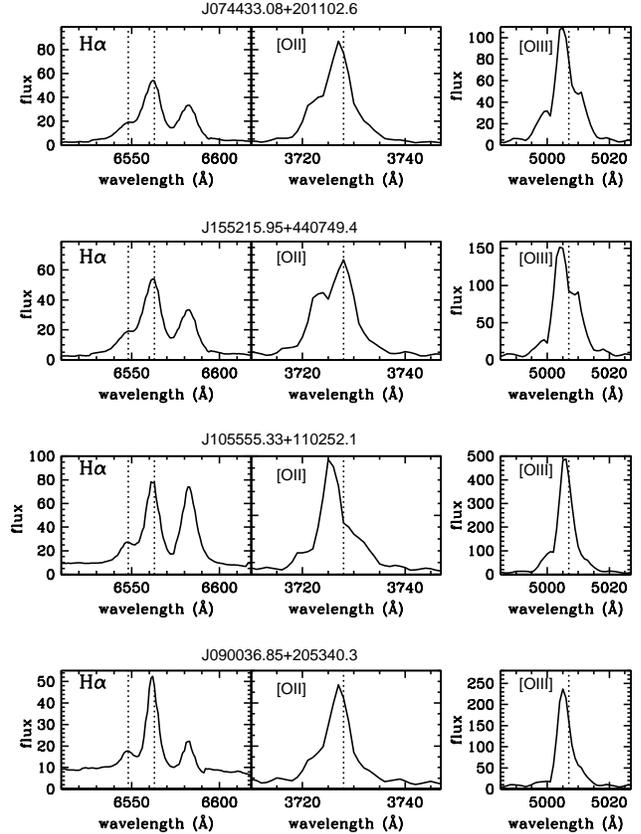}
\caption{ Four examples of nearby AGN with [OIII] line luminosities greater than
$10^9 L_{\odot} $ from the sample of Kauffmann (2018) with clear signatures
of outflowing gas. We plot the spectra in the vicinity of the H$\alpha$,
[OII]$\lambda$3727,3729 and [OIII]$\lambda$5007 lines. The vertical dotted
lines in the left panels  show the expected  location of H$\alpha$ and [NIII]$\lambda$6548
with no systemic shift. In the middle and right panels, we mark 
$\lambda$= 3728\AA and $\lambda$= 5007\AA.   
\label{models}}
\end{figure}

\begin{figure}
\includegraphics[width=90mm]{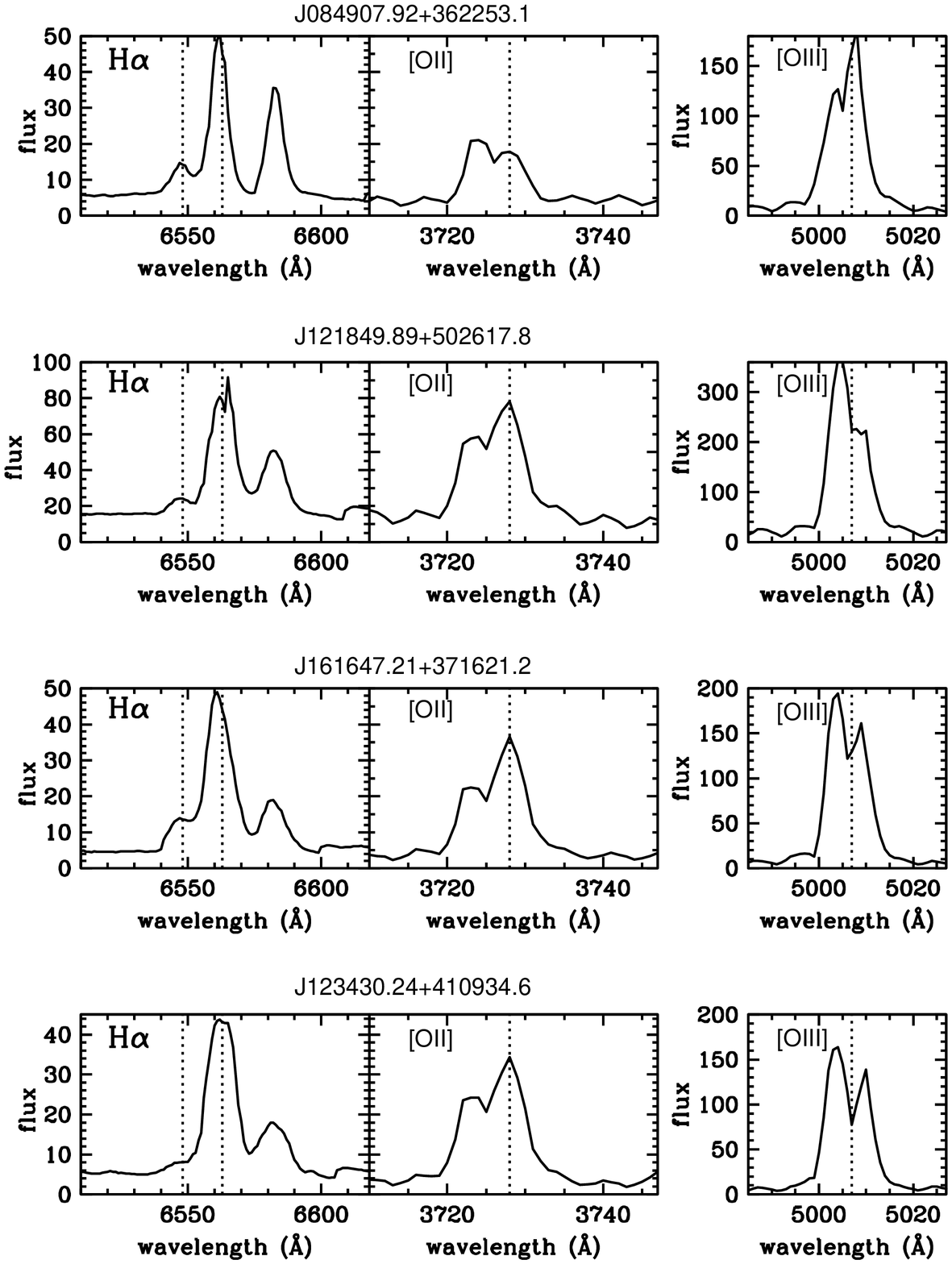}
\caption{ Four examples of nearby AGN with [OIII] line luminosities greater than
$10^9 L_{\odot} $ from the sample of Kauffmann (2018) with double-peaked 
narrow line profiles. We plot the spectra in the vicinity of the H$\alpha$,
[OII]$\lambda$3727,3729 and [OIII]$\lambda$5007 lines.  
The vertical dotted lines are the same as in Figure 8.
\label{models}}
\end{figure}

Interestingly, in our 4 double-peaked sources, the right peak is always centered
on the systemic velocity in the case of the [OII] line. For the [OIII] line, the
minimum between the two peaks is centered on the systemic velocity. Double peaks
are not seen in the H$\alpha$ line, except for one case.  These systematic
differences between different emission line tracers would appear to disfavour
the binary black hole hypothesis and favour a scenario where complex radiative
transfer effects within a single structure give rise to double peaked line
morphologies. For the stacking analysis presented in the next section, our
results for individual systems indicate that broad components at the {\em base}
of a central, more narrow emission line, are likely to be indicative of outflows
of ionized gas.

\section {Line component-fitting results for higher redshift stacked spectra} A secondary broad
component is detected in all the stacks for the [OIII]$\lambda$4959,
[OIII]$\lambda$5007 and H$\alpha$ lines, and in some stacks for the
[OII]$\lambda$3727,3729 doublet.  A summary of the properties of the broad
components is presented in Table 1.  We list the fraction of the total flux
contained in the broad component and the velocity widths of the narrow and broad
components for each stack.  The broad components of the [OII] and [OIII] lines
have velocity widths of around 500 km/s and contain about a third of the total
line flux. We have not yet implemented a formal computation of the
errors in these measurements, but examination of the variation between different
stacks indicate that the uncertainties are around 50 km/s for
the velocity width and 0.05 for the fraction of the flux in the broad component.   
 There is no clear trend in either the velocity width or the broad
component fraction with stellar age or with [OIII] line luminosity of the
galaxies in the stack.  We attribute the fact that the broad [OII] component is
not always clearly detected to the fact that the line is a doublet. Also, as
shown in the previous section, some sources will be double-peaked, leading to a
narrow component in the stack with larger line width. In Table 1, we see that
the line width of the narrow component of [OII] is 230-260 km/s compared to
190-210 km/s for the [OIII] line. This means that the contribution from
outflowing gas is harder to detect after stacking. The broad components of the
H$\alpha$ lines in the stacks also contain a third of the total line flux, but
are considerably more extended that those of [OII] or [OIII], with velocity
widths of 750-950 km/s. Once again, there is no clear trend in the widths of
these broad components with stellar age or [OIII] line luminosity.

\begin{table*}
  \begin{center}
    \caption{Table of line component quantities. The three quantities listed in each column are, 1) fraction of flux in
broad compoent, 2) $\sigma$ of narrow component (km/s), 3) $\sigma$ of broad component (km/s). ND means
non-detection of the broad component.} 
    \label{tab:table1}
    \begin{tabular}{l|l|l|l|l|l} 
      \textbf {log L[OIII]} & log age(Gyr)& \textbf{OII$\lambda$3727} & \textbf{OIII$\lambda$4959}& 
        \textbf{OIII$\lambda$5007} & \textbf{H$\alpha\lambda$6563}\\
      \hline
8.5 & 8.0   &             ND     &          0.33007; 199.88; 502.86 & 0.34719; 209.79; 557.97 & 0.34075; 241.33; 866.82\\   
8.5 & 8.4   &     0.17494; 224.91; 633.39 & 0.32681; 186.00; 467.96 & 0.34723; 197.94; 526.53 & 0.41579; 182.60; 948.53\\ 
8.5 & 8.8   &             ND     &          0.32078; 190.97; 457.97 & 0.34723; 197.94; 526.53 & 0.31070; 301.12; 920.78\\     
8.5 & 9.2   &             ND     &          0.32019; 196.80; 471.88 & 0.31954; 202.52; 475.54 & 0.34829; 245.36; 925.00\\      
      \hline
9.0 & 8.0   &     0.28912; 235.64; 479.79 & 0.33986; 188.18; 495.71 & 0.34053; 197.95; 511.12 & 0.32350; 180.60; 804.34\\ 
9.0 & 8.4   &     0.29682; 233.81; 494.28 & 0.32910; 204.49; 514.51 & 0.34052; 202.52; 522.96 & 0.30507; 263.61; 759.21\\ 
9.0 & 8.8   &     0.28125; 232.45; 455.11 & 0.32971; 197.91; 497.96 & 0.43324; 207.10; 502.47 & 0.29143; 302.80; 818.22\\      
9.0 & 9.2   &             ND              & 0.29593; 193.73; 418.78 & 0.26940; 202.88; 492.20 & 0.33236; 252.34; 861.52\\      
      \hline
9.5 & 8.0   &     0.27305; 257.81; 484.61 & 0.30817; 221.44; 504.90 & 0.31222; 230.02; 522.14 & 0.33023; 270.15; 922.21\\     
9.5 & 8.4   &     0.26478; 250.48; 451.39 & 0.30803; 224.52; 511.94 & 0.29461; 220.66; 500.89 & 0.30071; 319.04; 918.85\\     
9.5 & 8.8   &             ND              & 0.33058; 192.15; 483.51 & 0.32668; 201.02; 487.62 & 0.34838; 256.52; 967.16\\      
      \hline
    \end{tabular}
  \end{center}
\end{table*}

In Figures 10, 11 and 12 we illustrate our fits to the lines in the stacked
spectra for three different cases: a) the [NeIII]$\lambda$3869 line, for which
only one component is detected in all cases (Figure 10), b) the
[OIII]$\lambda$5007 line, for which a broad component is always detected (Figure
11), c) the H$\alpha$ and [NII]$\lambda$6584 line complex (Figure 12). We show
results for the $\log$ L[OIII]=9.0-9.5 bin for three different ranges in stellar
age. In the top row, we show results for the full stacks, in the middle row for
stacks of radio-detected sources with the same [OIII] luminosity and age, and in
the bottom row for stacks of WISE-detected sources. In Figure 10 and 11, the
solid black curves show the emission lines in the stacked spectra  and the solid
red curves the final fit. In the case of a two component fit, the broad
component is shown as a dashed magenta curve, while the narrow component is
shown as a dashed red curve.  The cyan shaded region indicates the errors on the
flux in the stacked spectrum, estimated through a bootstrap resampling technique.

Examining Figures 10 and 11, we see that the [NeIII] and [OIII] profiles for the
radio and WISE-detected subsamples are very similar to those of the full sample
at fixed luminosity and stellar age. The radio-detected subsample shows evidence
for complex structure in the narrow component, particularly at young stellar
ages. We do not find any systematic quantitative differences between the broad
components of the radio-loud and WISE-detected subsamples, showing that the
large-scale ionized gas outflows in these systems are identical to those in the
luminous AGN population as a whole. This result does not support claims that
that radio jets are critical feedback mechanisms acting on the ionized gas in
massive galaxies (Jarvis et al 2019).

It is also worth noting that the non-detection of a broad
component of the [NeIII]$\lambda$3869 line is not likely to be a consequence
of the fact that this line is weaker in flux than  [OIII]$\lambda$5007. 
A broad component of the [NeIII] line of one third of the
total flux would be of the same magnitude  as other well-detected lines in
the stacks. The more likely explanation is that the outflowing gas traced
by [NeIII], which has higher ionization
potential than [OIII], is confined closer to the nucleus of the galaxy and
cannot be distinguished from the narrow-line region gas. 
As we will now show,  H$\alpha$, which traces the coolest and densest gas,
exhibits the most extended broad component.

\begin{figure}
\includegraphics[width=90mm]{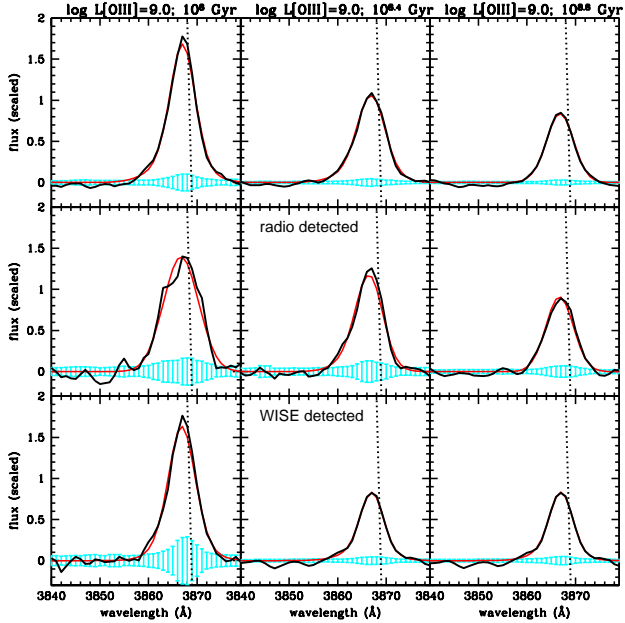}
\caption{Illustration of fits to the [NeIII]$\lambda$3869 line for stacked spectra
from our high redshift sample
in the L[OIII]=$10^9$ $L_{\odot}$ line luminosity bin at three different stellar ages.  
The solid back curve shows the stacked spectra and the red curve is the
single Gaussian fit. The middle and lower panels compare results for stacked spectra
of radio and WISE detected galaxies with the same [OIII] luminosities and 
stellar ages as those in the full sample.
The cyan shaded region indicates the errors on the
flux in the stacked spectrum, estimated through a bootstrap technique.
The vertical dotted line shows  $\lambda$= 3869\AA
\label{models}}
\end{figure}

\begin{figure}
\includegraphics[width=90mm]{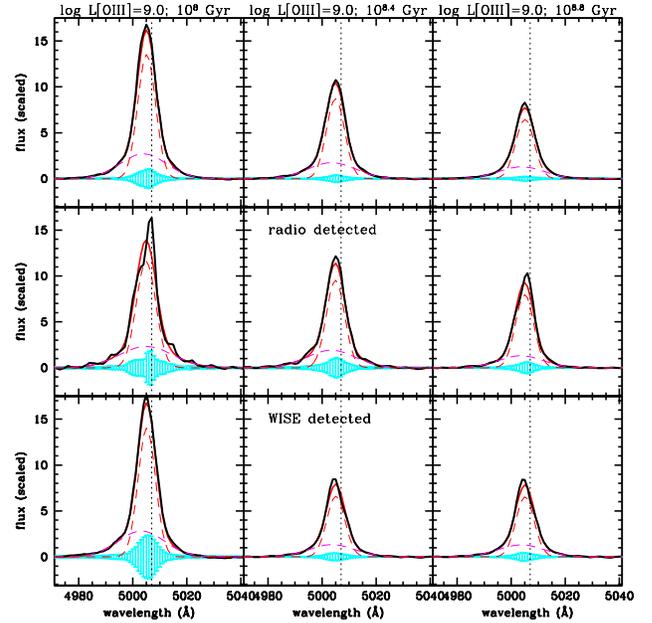}
\caption{Illustration of fits to the [OIII]$\lambda$5007 line for stacked spectra
in the $\log$L[OIII]=9 line luminosity bin at three different stellar ages.  
The solid back curve shows the stacked spectra and the solid red curve is the
double  Gaussian fit. The broad
component is shown as a dashed magenta curve, while the narrow component is
shown as a dashed red curve. The middle and lower panels compare results for stacked spectra
of radio and WISE detected galaxies with the same [OIII] luminosities and 
stellar ages as those in the full sample.
The cyan shaded region indicates the errors on the
flux in the stacked spectrum, estimated through a bootstrap technique.
The vertical dotted line shows  $\lambda$= 5007\AA
\label{models}}
\end{figure}

Our fits to the H$\alpha$/[NII] complex are carried out in a series of steps.
In Figure 12, the truncation in the  H$\alpha$ line redwards of the peak delineates
the wavelength region over which the fit is performed. Once again, the solid red curve
shows the full fit, while the dashed red and magenta curves show contributions
from the  narrow and broad components, respectively. The solid magenta curve
shows the best single Gaussian fit to the [NII]$\lambda$6584 line after the
broad component of  the H$\alpha$ line is subtracted. As can be seen, two
Gaussians for H$\alpha$ plus a single Gaussian for [NII] provide a
good fit to the line complex as a whole. A net blue side excess is visible in
most of the stacks. The fact that a net blue side excess is seen only for
H$\alpha$ stack and not for the [OIII] stack is likely explained by the fact
that H$\alpha$ traces cooler, denser gas whose emission is more affected by dust
obscuration.

\begin{figure*}
\includegraphics[width=143mm]{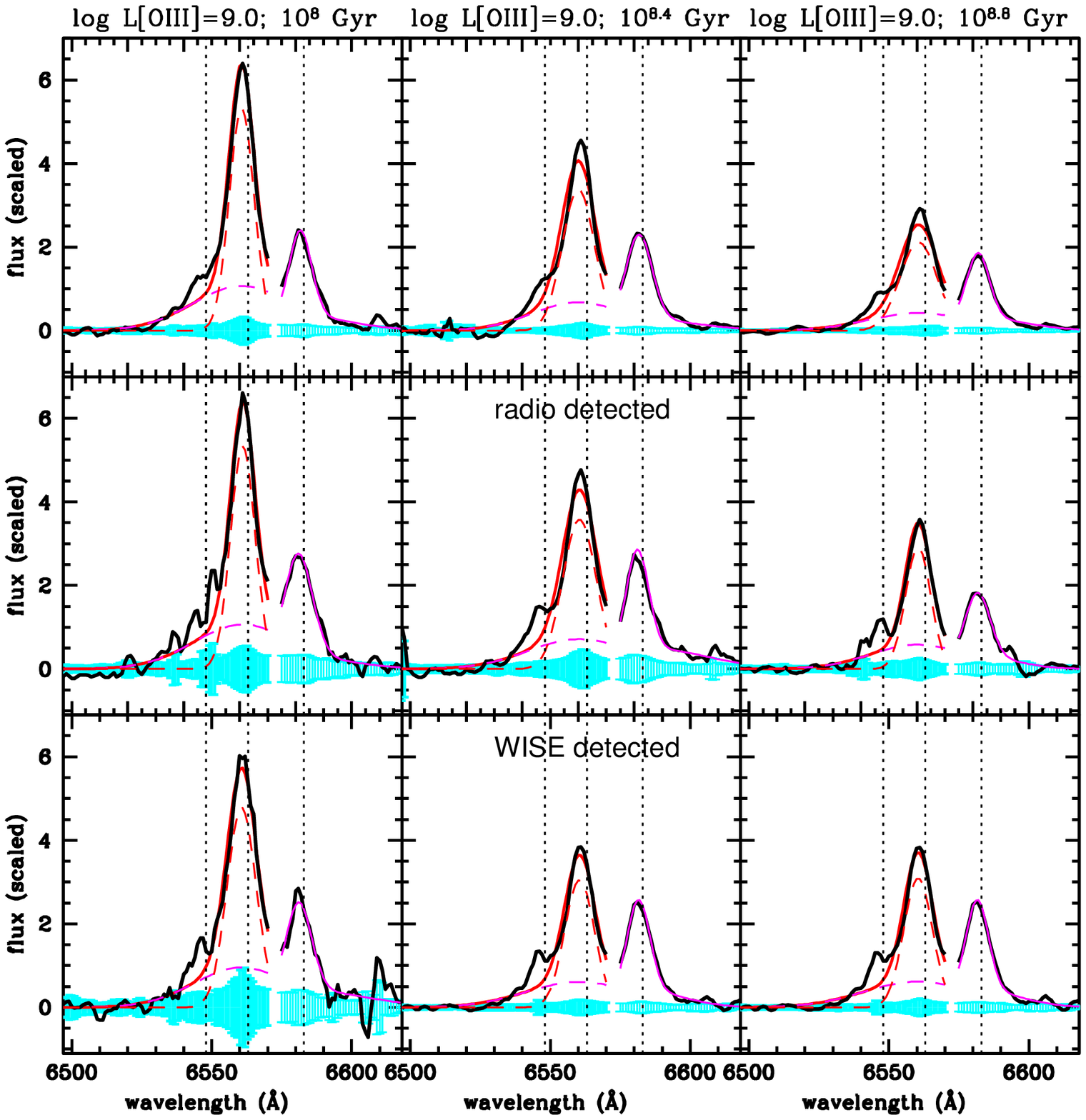}
\caption{Illustration of fits to the H$\alpha$  line for stacked spectra
in the $\log$L[OIII]=9 line luminosity bin at three different stellar ages.  
The solid back curve shows the stacked spectra. solid  The red curve is the
double  Gaussian fit, with the broad
component shown as a dashed magenta curve and the narrow component 
as a dashed red curve. The solid magenta curve shows the single Gaussian 
fit to [NII]$\lambda$6583.  The middle and lower panels compare results for stacked spectra
of radio and WISE detected galaxies with the same [OIII] luminosities and 
stellar ages as those in the full sample.
The cyan shaded region indicates the errors on the
flux in the stacked spectrum, estimated through a bootstrap technique.
The three vertical dotted lines mark the $\lambda$= 6548\AA [NII] line
and the rest-frame wavelengths of H$\alpha$ and [NII]$\lambda$6583.
\label{models}}
\end{figure*}

A number of quantitative measures of the H$\alpha$ blue wing excess are provided
in Table 2 for each stacked spectrum. We define the blue excess region as the
contiguous wavelength region over which the measured line flux remains 1$\sigma$
above the flux given by the best-fit, two-component Gaussian model.  We tabulate
the minimum wavelength of this region, the corresponding velocity separation of
the minimum with respect to the systemic velocity, and the fraction of the broad
line component flux that is in the blue excess. The latter is obtained by
subtracting the best-fit model flux from the measured flux, summed over the 
wavelength region that spans the blue excess. The typical velocity
separations are around $\sim$ 1000 km/s, with a scatter of $\sim 200$  km/s
between different stacks in a given luminosity bin. The scatter between the measurements of the blue
excess fraction between different stacks in the same luminosity bin is $\sim 0.02$.
There are indications that lower luminoisty AGN
have larger H$\alpha$  blue excess fractions, even though the maximum blue shift
of this component is not larger than for the higher luminosity AGN. 
We note that we have not attempted to subtract the contribution from [NII]$\lambda$6548.
This line is so heavily blended with the H$\alpha$ line that any attempt to do
so would need to assume that there is a fixed relation with the [NII]$\lambda$ 6583 line.
The blue excess fractions should thus be taken as upper limits and attaention given
to relative trends rather than absolute values for this quantity.

\begin{table*}
  \begin{center}
    \caption{Table of quantities pertaining to the blue excess seen in the H$\alpha$ line.} 
    \label{tab:table1}
    \begin{tabular}{l|l|l|l|l} 
      \textbf {log L[OIII]} & log age(Gyr)& $\lambda$(blue) \AA  & V(blueshift) km/s & 
        F(blue)/F(H$\alpha$)\\
      \hline
8.5    &      8.0      &          6544   &      871     &          0.077 \\                      
8.5    &      8.4      &          6536   &     1239     &          0.081 \\              
8.5    &      8.8      &          6536   &     1239     &          0.118 \\                
8.5    &      9.2      &          6536   &     1239     &          0.071 \\               
      \hline
9.0    &      8.0      &          6540   &     1055     &          0.0520 \\             
9.0    &      8.4      &          6530   &     1516     &          0.0465 \\             
9.0    &      8.8      &          6537   &     1193     &          0.0598 \\            
9.0    &      9.2      &          6437   &     1193     &          0.0973 \\            
      \hline
9.5    &      8.0      &          6538   &     1147     &          0.0427 \\              
9.5    &      8.4      &          6531   &     1469     &          0.0367 \\              
9.5    &      8.8      &          6538   &     1147     &          0.0583 \\              
      \hline
    \end{tabular}
  \end{center}
\end{table*}

So far, we successfully identified broad components in stacked spectra, but we
have not found significant variation in the properties of this component with
AGN luminosity, stellar age, radio or dust emission properties.  Although the
galaxies in our sample span a rather restricted range in stellar mass, it is
still possible to split the sample into two well-separated  stellar mass ranges
for the most populated bins. 
We note that [OIII] is the highest S/N line in all the stacks. In
Figure 13, we show results for AGN with $\log$ L[OIII]=9.0-9.5 and stellar ages
$\log$ age =8.4.-8.8 . Results for stellar masses in the range $3 \times
10^{10}-10^{11} M_{\odot}$ are shown in the left panel and for masses greater
than  $3 \times 10^{11} M_{\odot}$ in the right panel. As can be seen from Figure 1,
these mass ranges lie on the tails of our stellar mass distribution and 
we are thus only able to show results for our most populated [OIII] luminosity
and stellar age bin.   We find a 40\%
difference in the width of the broad component for the two subsamples.  The fact
that the width of the broad component is more dependent on the
potential well depth of the host galaxy than it is on stellar age, [OIII], radio
or mid-IR luminosity, suggests that the galaxy halo plays a crucial role in
regulating the kinematics of the outflowing ionized gas, most likely via
gravitational confinement and processes such as shock-heating and virialization.
Alternatively, it could also mean that the mass of the black hole, which is
well-correlated with the stellar mass of the galaxy, is the determining factor.
We will come back to these issues in the next section.

\begin{figure}
\includegraphics[width=90mm]{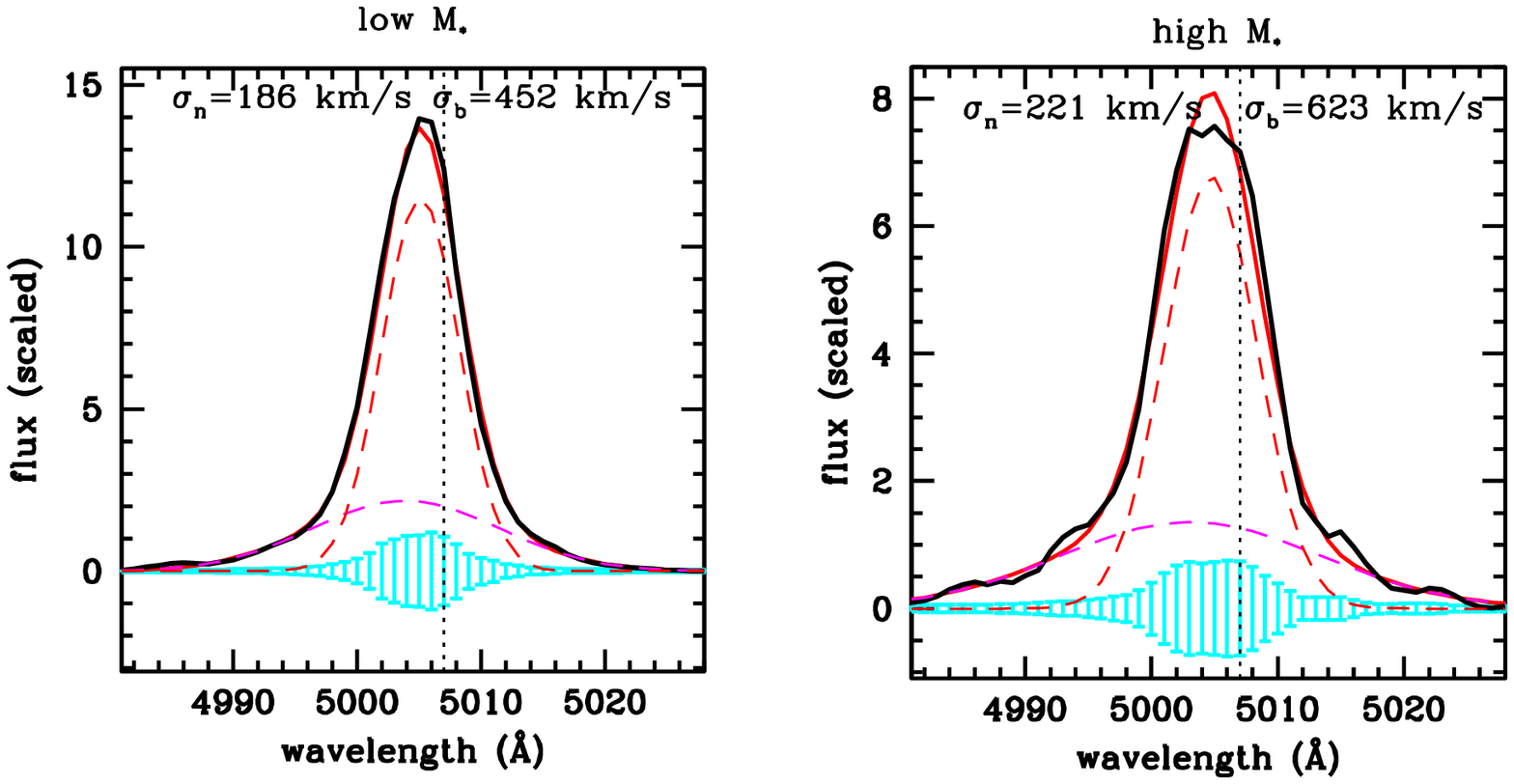}
\caption{Illustration of fits to the [OIII]$\lambda$5007 line for stacked spectra
in the $\log$L[OIII]=9 line luminosity bin and stellar ages 
$\log$ age =8.4.-8.8 . Results for stellar masses in the range $3 \times
10^{10}-10^{11} M_{\odot}$ are shown in the left panel and for masses greater
than  $3 \times 10^{11} M_{\odot}$ in the right panel. The figure format
is the same as in Figure 11.  
\label{models}}
\end{figure}

\section {Summary and conclusions} 
We have carried out a systematic analysis of
the emission line properties of Type II AGN at redshifts between 0.4-0.8  with
[OIII] luminosities greater than $3 \times 10^8 L_{\odot}$, i.e systems with
luminosities characteristic of the Type II quasars first identified in
population studies by Zakamska at at (2003).  These luminous AGN are drawn from
the  CMASS sample of galaxies that were studied as part of the Baryon
Oscillation Spectroscopic Survey (BOSS) survey of galaxies with stellar masses
greater than $10^{11} M_{\odot}$, and comprise 0.5\% of the total population of
galaxies at these redshifts. Individual spectra have low S/N, so our analysis is
carried out on stacked spectra in bins of [OIII] luminosity and stellar age.

Our main findings are the following: \begin{itemize} 
\item 
The emission line ratios of the stacks  are 
remarkably well-fit with simple uniform-density
photo-ionization models with power-law ionizing spectra and metallicities
between solar and twice solar. The only line ratio that is found to be
significantly anomalous is [HeII]$\lambda 4686$/H$\beta$, but only in the most
luminous AGN at a stellar age of a few times $10^8$ years. The stellar continuum
in these objects shows very prominent post-starburst features. 
\item In the stacks, a number of emission lines  
are found to have distinct, well-detected
broad components requiring a double Gaussian rather than a single Gaussian fit.
These are: [OIII]$\lambda$4959, [OIII]$\lambda$5007, [OII]$\lambda$3727,3729 and
H$\alpha$$\lambda$6563.  Higher ionization potential  lines such as [NeIII] and
[NeV] are detected in the stacks, but well-fit by single Gaussians with small
net blue-shifts.  
\item A net blue-side excess after double Gaussian fitting is
detected in the H$\alpha$ stack but not in the [OII] or [OIII] stacks. 
\item In the stacks, the detected broad components always contain a third of the total
line flux.  
\item The width of the broad component is $\sim 500-700$ km/s for
the [OII] and [OIII] lines and $\sim 750-950$ km/s for H$\alpha$.  
\item The fraction of the flux in the broad component and its width are independent of
[OIII] luminosity, stellar age, radio and mid-IR luminosity.  
\item The only parameter we have identified in this study 
that appears to influence the width
of the broad component is the stellar mass of the galaxy $M_*$.
\end{itemize}

The strongest evidence that the broad components are related to 
outflowing gas on large scales rather than gas in the accretion disk near  the
black hole (so called broad-line region (BLR) gas) comes from the fact that the
stellar continuum in all the stacks are well fit using linear combinations
of simple stellar populations. We performed the experiment of allowing for
a featureless power-law component to model a possible contribution from the
accretion disk and this did not improve our fits.  We also checked for
coronal Fe lines and did not find any clear detections. Finally, the lack of
a clear broad component in the higher ionization potential [NeIII] and [NeV]
lines also disfavours the hypothesis that we are seeing BLR gas in these systems.

Stacking together many different galaxy spectra has the disadvantage of  washing
out complexity. As we have discussed in Section 1, IFU studies reveal complex,
multi-phase outflows with significant spatial and velocity structure in
individual systems. Nevertheless, the hope is that stacking may also yield a
number of simple conclusions that provide some degree of physical insight.

One conclusion that has emerged from multi-phase studies of individual systems,
such as the study of the molecular and  ionized components of the AGN-driven
outflow in zC400528, a massive,  main sequence galaxy at z=2.3 in the process of
quenching (Herrera-Camus et al 2019), is that although the molecular outflow
dominates the mass and energy budget, most of the gas does not reach very high
velocities.  Our study confirms the result shown in Figure 3 of this paper that
H$\alpha$ effectively probes gas located in the tails of the velocity
distribution ($\pm 500-1000$ km/s), well-separated from  the systemic velocity
of the galaxy.  It is interesting that these high velocity tails are not seen in
the higher-ionization lines, suggesting that there may be a ``sweet spot''
in gas density and temperature for probing such material and that  H$\alpha$  
may be the most effective tracer.  

Another conclusion in  need of an explanation is the lack of correlation between
the properties of the broad ionized gas component and properties such as AGN
luminosity, mean stellar age, and whether or not the galaxy is detected at radio
or mid-IR wavelengths.  For AGN molecular outflows, Veilleux et al (2013) find that
while  the outflow  velocities  show  no  statistically  significant  dependence
on star formation rate, they are distinctly more blue-shifted among systems with
larger AGN luminosity. Perna et al (2017) find a weak correlation between
outflow velocity and AGN luminosity for a sample of X-ray selected AGN.  
We caution that our study is
restricted to massive galaxies ($\log M_*$ > 11), which reside in 
dark matter halos that are massive enough  to contain quasistatic hot gas atmospheres.
In such a situation, the AGN-driven wind may be largely confined
by the pre-existing gas in the halo and the ionized gas kinematics
on large scales may partially reflect virial motions of the halo gas.

Recent work by  Nelson et al (2019) on outflows in simulated
galaxies has emphasized how the relative simplicity of model inputs (and
scalings) at the injection scale produces complex behavior at galactic and halo
scales. In their simulations, however, the outflow velocities at injection
increase strongly  with  $M_*$, and those measured on galactic scales increase
only slightly more weakly, suggesting that  some memory of the physical
processes at injection should be retained. We caution that in these simulations
there is no attempt to link outflow properties with the energetic output from
AGN in the form of radiation, so it would be difficult to make robust
comparisons with the data.  Observationally, separating the physics at the wind
injection scale from the physics on halo scales  will likely require
well-resolved studies of large samples of galaxies spanning a wide range in
stellar mass and also multi-phase information about the gas.
This is  certainly a very significant observational challenge for the future.

\noindent
{\bf Acknowledgments}\\
We thank Dylan Nelson for useful comments.

%===================================

\end{document}